# Seven Clusters In Genomic Triplet Distributions

Alexander N. Gorban[2,3], Andrei Yu. Zinovyev[1*], Tatyana G. Popova[2]

[1]*Institut des Hautes Études Scientifiques, Bures-sur-Yvette, France*
[2]*Institute of Computational Modeling, Russian Academy of Science*
[3]*Institute of polymer physics, ETH, Switzerland*

**ABSTRACT:** In several recent papers new gene-detection algorithms were proposed for detecting protein-coding regions without requiring learning dataset of already known genes. The fact that unsupervised gene-detection is possible closely connected to existence of a cluster structure in oligomer frequency distributions. In this paper we study cluster structure of several genomes in the space of their triplet frequencies, using pure data exploration strategy. Several complete genomic sequences were analyzed, using visualization of tables of triplet frequencies in a sliding window. The distribution of 64-dimensional vectors of triplet frequencies displays a well-detectable cluster structure. The structure was found to consist of seven clusters, corresponding to protein-coding information in three possible phases in one of the two complementary strands and in the non-coding regions with high accuracy (higher than 90% on the nucleotide level). Visualizing and understanding the structure allows to analyze effectively performance of different gene-prediction tools. Since the method does not require extraction of ORFs, it can be applied even for unassembled genomes.

## INTRODUCTION

With few exceptions, almost all commonly used gene-finding programs employ a learning dataset for tuning parameters of a learning rule. In several recent papers new algorithms were proposed for detecting coding regions without requiring learning dataset of already known genes. In Bernaola (2000) the authors proposed a method developed for unsupervised segmentation of whole DNA texts corresponding to the division on coding and non-coding regions. In Audic and Claverie (1998) the authors proposed a clustering procedure which uses all available annotated genomic data for its calibration. This iterative procedure uses genomic sequences to adjust parameters (that are initialized by randomly partitioning a number of small subsequences) of several probabilistic sequence models. The algorithm converges fast and gives accuracy up to 90%. In Baldi (2000) it was explained that this algorithm is essentially a form of the expectation maximization algorithm applied to the corresponding probabilistic mixture model.

The fact that unsupervised gene-detection is possible and effective (and, to lesser extent, supervised learning as well) closely connected to existence of a cluster structure in oligomer distributions. Implicitly existence of this structure is known (see, for example, Borodovsky, 1993) and is widely used in gene-recognition, but was never visualized and studied in terms of data exploration strategies. Visual representations of the structure allows deeper understanding of its properties and can serve as useful display for analyzing characteristics of existing gene-finders.

In this paper we use a method of data visualization to explore the space of frequencies of triplets in a sliding window. We demonstrate and analyze the structure of datasets used both for supervised and unsupervised learning.

In this work we do not have purpose to invent principally new gene-prediction tools, but rather provide illustrations of the datasets that are utilized by existing prokaryotic gene-finders. Just as one quantative illustration, we propose a simple clustering method for detecting coding regions and assess it, using standard measures. Its performance turns out to be similar as of the methods mentioned above, as well as of modern prokaryotic gene-finders. In addition, even this simplified method, which does not take into account many important details, can have practical value when application of ORF extraction strategies is not possible (in the case of unassembled genomes, for example, when only chunks of 200-500bp length are available).

*Corresponding author E-mail: zinovyev@ihes.fr.

## Methods

Let us denote a codon frequency distribution as $f_{ijk}$ , where *i,j,k* are in the {A,C,G,T} set, i.e., for example, $f_{ACG}$ is equal to the relative frequency of the ACG codon in a given coding region. One can introduce such natural operations over frequency distribution as *phase shifts* $P^{(1)}$ , $P^{(2)}$ and *complementary reversion* $C_R$:

$$P^{(1)} f_{ijk} \equiv \sum_{l,m,n} f_{lij} f_{kmn} \, , \; P^{(2)} f_{ijk} \equiv \sum_{l,m,n} f_{lmi} f_{jkn} \, , \; \hat{f}_{ijk} \equiv C^R f_{ijk} \equiv f_{\hat{k}\hat{j}\hat{i}} ,$$

where $\hat{i}$ is complementary to *i, i.e.,* $\hat{A} = T$ , $\hat{C} = G$, etc.

The phase-shift operator $P^{(n)}$ calculates the new triplet distribution, but now counted with a frame-shift on *n* positions, in the hypothesis that no correlations exist in codon order. Complementary reversion constructs the distribution of codons from a coding region in the complementary strand, but counted in the forward strand (“shadow” codon usage).

The phase-shift operator $P^{(n)}$ calculates a new triplet distribution, counted with a frame-shift on *n* positions, in the hypothesis that no correlations exist in the codon order. Complementary reversion constructs the distribution of triplets in the complementary strand, counted in the direct strand ("shadow" codon usage).

Let us introduce the distance between two distributions as

$$\left\| f_{ijk} - g_{ijk} \right\| = \sum_{ijk} \left| f_{ijk} - g_{ijk} \right| .$$

It is then natural to expect that the problem of gene recognition can be solved if one of the numbers, $\left\| f_{ijk} - P^{(1)} f_{ijk} \right\| , \left\| f_{ijk} - P^{(2)} f_{ijk} \right\|$ is large enough. It follows from that remark that after a large number of insertion and deletion operations of one base-pair at a time, we would have

$$\left\| f_{ijk} - P^{(1)} f_{ijk} \right\| \approx 0 \, , \left\| f_{ijk} - P^{(2)} f_{ijk} \right\| \approx 0 \, .$$

This is expected to happen in non-coding regions, where frameshifts do not necessarily lead to misreading of genetic material, and eventially happen due to mutations.

Let us introduce a measure of how far $f_{ijk}$ is from the shifted distributions:

$$CP = \max\left( \left\| f_{ijk} - P^{(1)} f_{ijk} \right\| , \left\| f_{ijk} - P^{(2)} f_{ijk} \right\| \right)$$

Real (counted directly from genetic texts) triplet frequency distributions in the first and the second phases will be denoted as $f_{ijk}^{(1)}$, $f_{ijk}^{(2)}$, $\hat{f}_{ijk}^{(1)}$, $\hat{f}_{ijk}^{(2)}$. Let us introduce the term *"codon correlation contribution measure"* as the average distance between real and calculated distributions:

$$CC = \frac{1}{2} \left( \left\| P^{(1)} f_{ijk} - f_{ijk}^{(1)} \right\| + \left\| P^{(2)} f_{ijk} - f_{ijk}^{(2)} \right\| \right).$$

We constructed datasets of triplet frequencies for several real genomes and for several model genetic sequences, as follows:

1. Only direct strands of genomes are used for counting triplets;

2. Every $p$ positions in a sequence, we open a window $(x\text{-}W/2,x+W/2)$ of size $W$ and centered at position $x$; here $p>>1$ is used to reduce sample size of the resulting dataset, otherwise one has to deal with sample size of millions of points, while introducing $p$ the sample size s $[L/p]$ points, where $L$ is the entire length of the sequence;

3. Every window, starting from the first base-pair, is divided into $W/3$ non-overlapping triplets, and the frequencies of all triplets $f_{ijk}$ are calculated;

4. The dataset consists of $N = [L/p]$ points; every data point $X_i=\{x_{is}\}$ corresponds to one window and has 64 coordinates, corresponding to the frequencies of all possible triplets s = 1,…,64. Resulting datasets were used both for visualization and clustering.

The standard centering and normalization on unit dispersion procedure is then applied, i.e., $\tilde{x}_{is} = \frac{x_{is} - m_s}{\sigma_s}$, where $\tilde{x}_{is}$ is the value of the $s$th coordinate of the $i$th point after normalization, and $m_s$ is the mean value of the sth coordinate, $\sigma_s$ is the standard deviation of the $s$th coordinate.

Then we applied principal components algorithm in order to visualize these 64-dimension datasets projected onto the 3-dimensional linear manifold spanned by the first three principal vectors of the distribution. It is known that projection onto this manifold is only as informative as higher the value of $v^{(3)} = D^{(3)}/D$, where $D$ is the dispersion of the dataset, calculated in 64-dimensional data-space and $D^{(3)}$ is the analogous quantity calculated after projecting the vectors into the 3-dimensional space. In practice, even if the value of $v^{(3)}$ is not high enough (say, it equals 0.1-0.3), we may still try to visualize the dataset, in the hope of being able to pick up qualitative "signals" of presence of patterns in the data distribution, as well as to visually represent the dataset.

## Results

Figure 1 presents several distributions calculated for 4 genetic texts.

In addition to the distribution itself, we introduced two triangles, formed by the points $f_{ijk}, P^{(1)} f_{ijk}, P^{(2)} f_{ijk}$ and $\hat{f}_{ijk}, P^{(1)} \hat{f}_{ijk}, P^{(2)} \hat{f}_{ijk}$, into the figures. The large spheres correspond to the points $f_{ijk}$ and $\hat{f}_{ijk}$, where $f_{ijk}$ was calculated from the genome's known annotation. Data-points have different shapes and colors, according to whether they are coding or non-coding in one of the two strands.

The explanation of the structures is quite evident: Coding information from windows in the direct strand has one of three possible phase shifts. Since this phase shift is not known in advance, approximately one-third of the windows fall into the vicinity of the point that corresponds to the $f_{ijk}$ (0-shift), one-third are close to the $f^{(1)}_{ijk}$ (1-shift), and the last third are close to the $f^{(2)}_{ijk}$ (2-shift). This is also true for the reverse strand, but with the centers corresponding to complementary distributions.

The plots shown at fig.1 are two-dimensional projections of 3D scatters. The first 2 distributions show very clear separation on seven clusters; no surprise, that in these cases unsupervised gene-prediction gives both high specificity and sensitivity. The distribution of triples in *s.cerevisiae, chromosome IV* forms seven clusters as well; though they are not clearly seen on 2D-pictures, because two "phase triangles", projected into the principal subspace are positioned on two parallel planes, perpendicular to the direction of the third principal component. The situation is worse in case of *p.wickerhhamii* mitochondrion genome. In this case distributions of triplets in the direct and reverse strands indeed overlap. Fig.1 shows that this is not simply because the triplet distributions in direct and reverse strands are similar, but 0-phase of the distribution of triplets in the direct strand overlaps with 1-phase of the distribution in the reverse strand, and so on. One can predict in this case that gene-recognition procedures will often mix genes in the direct and reverse strands, though ORF-strategies can probably resolve this conflict.

One can see from the pictures that the centers of phase-shifted distributions are close enough to the calculated points, demonstrating absence of significant correlations in the order of codons. Indeed, the calculated values of *CC* are not high (see Table 1, *CC* column.) It means that in real texts correlation between subsequent codons is much less then the "inter-phase" difference.

### *Clusterization*

Having in mind the visual representation of the distribution of data-points, it is possible to propose a natural way of segmenting sequences into regions that are homogeneous with respect to the coding phase. One would expect that regions with the same coding phase correspond to protein-coding regions.

We must underline, that in this work we do not have purpose to invent a principally new gene-prediction tool, but rather provide illustrations on the datasets that are utilized by existing prokaryotic gene-finders. The gene-finding method we describe below is just an illustration to the "seven-clusters" structure of the triplet distributions, and it is intentionally made as simple as it can be. Nevertheless, it can be used in the situations where application of ORF strategies is not possible, giving rather good performance characteristics.

Trying to be as simple as we can, we make use of one of the simplest clusterization strategies, namely unsupervised K-Means clustering. The clustering is accomplished in the 64-dimensional space and the positions of all seven clusters are identified. This is the result of the learning step of the procedure. Also one must store the coefficients which were used for normalization (on unit dispersion) of the dataset.

After that step, during classification of all sequence positions, one must assign the corresponding cluster label along the whole sequence. This can be done by opening a window of width $W$ at every 3rd position $x$ of the sequence (x-W/2,x+W/2) and calculating non-overlapping triplet distribution inside this window. This distribution, after normalization with the coefficients stored at the previous step, corresponds to a point in 64-dimnesional space. Then the closest cluster is determined in this space. If it is the central cluster, that point is assigned to be non-coding; otherwise it is assigned to one of three possible coding phases. These basic steps of the method are presented graphically on fig.2.

To evaluate the ability of this procedure to differentiate between "coding" and "non-coding" base-pairs, we used base-level sensitivity and specificity of exon recognition, the measures which are commonly used in this case:

$$Sn = \frac{TP}{TP + FN}, Sp = \frac{TP}{TP + FP}$$

where $TP$ is the number of true-positives, i.e., coding bases predicted to be coding; $TN$ is the number of true-negatives, i.e., non-coding bases predicted to be non-coding; $FP$ is the number of false-positives, i.e., non-coding bases predicted to be coding, and $FN$ is the number of false-negatives, i.e., coding bases predicted to be non-coding.

The results are shown in the *Sn1* and *Sp1* columns of Table 1.

We must underline that the procedure is fully automated and does not require any human intervention. Neither genomic annotation nor ORF strategy is used. The procedure is completely based on opening sliding window every *p* positions in a text, and "blind" counting frequencies of non-overlapping triplets, startiing from the first base-pair. Visualization of datasets can be useful to evaluate how reliable prediction will be (measuring compactness of the clusters, for example) and to compare prediction with known information. The only parameter - window size - may be visually evaluated by comparing pictures of data constructed with various values of W (see the full version of the paper on the accompanying web-page.) In fact, the dependence of effectiveness on window-size is not strong over a rather long interval of W.

*Using known data*

In the method described above, the learning process used no information other than sequence itself; it was completely "unsupervised". One can also try to make use of some extra knowledge, as discussed in the next paragraph.

Studying a set of training examples (for example, following the strategy of GLIMMER, using long ORFs as a training set), it is possible to explicitly calculate the centers of all seven clusters. We have done this, using annotation of the analyzed genomes. First, half of the genes were used to calculate the centers, and the rest for accuracy testing. Using these seven vectors as centroids, we calculated new values for the sensitivity and specificity of gene recognition. They are shown in the *Sn2* and *Sp2* columns of Table 1. Here no clusterization is made at all. We provide these numbers only to show how unsupervised learning is close to the supervised classification based on heuristics and biological intuition (using long ORFs or homology search, for example).

*Our method and GLIMMER gene-finder*

Choosing GLIMMER [Salzberg et al., 1998, Delcher et al., 1999] for our analysis was dictated by our desire to use a gene-predictor that does not use any learning information, except that can be extracted from the genetic sequence itself. In GLIMMER, a learning dataset is formed by extracting long ORFs (usually longer than 500 bp) and then a variant of HMM-based predictor is used. Thus GLIMMER extrapolates the model of genetic sequence derived from the longest ORFs onto the shorter ORFs, which are the genes candidates. It is known that GLIMMER has a tendency to produce a lot of false-positive predictions. That version of GLIMMER that we used (version 2.02) did not have any model for non-coding regions. It was interesting to understand for us if and how many of GLIIMER false-positive predictions are due to this lack.

GLIMMER gene-finder uses some ORF strategy to detect potential genes. Because of this, we have to introduce simple rules for deciding if a given ORF is coding or non-coding, in our "seven-clusters" methodology. For every ORF, we calculate 64-dimensional vector of it's triplet frequencies and find the closest centroid in the triplet frequencies space (the positions of the centroids are calculated as it was described earlier). If the closest centroid is the one, which corresponds to the correct coding phase (let us denote it by P0), then this ORF is assigned to be coding. After this, from all such ORFs in P0 phase we filter out all ORFs that are too distant from the P0-centroid (the threshold is determined by an additional parameter), and all ORFs which are inside other ORFs in the P0 phase (it means that we take the longest ORF in the P0 phase).

To test this procedure, we analyzed output of GLIMMER gene-finder (using default settings), using the list of ORFs, produced by GLIMMER. Thus, we compare only effectiveness of the measures used, and not the details of ORF extraction strategy.

In the table 2 we show the results of this comparison, using existing annotations of the genomes in GenBank. One must understand that the annotations are far from being perfect and some part of the ORFs that we denoted as false positives in the GLIMMER prediction can be unknown putative genes (as it is claimed by the authors of GLIMMER). Nevertheless, we find significantly lower false-positive rate of our method comparing to the GLIMMER prediction. Analyzing this, in some genomes we found that a cluster structure exists in the distribution of false-positive GLIMMER predictions. On fig.3 visualization of GLIMMER predictions on the principal components plane is shown for Escherichia coli and Caulobacter crescentus for which GLIMMER produces many predictions of "additional genes". For example, our analysis shows that 62% of false-positives predictions for Escherichia coli and 80% of false positives for Caulobacter crescentus in the 64-dimensional space of triplet frequencies are closer to the centroid, which corresponds to the $C_R\, f_{ijk}$ distribution (C0-centroid), while only 2% of true-positive predictions for Caulobacter crescentus are close to the C0-centroid. Such discrepancy cannot be explained simply by "presence of unknown genes" but it is due to some effect of this HMM-based predictor, which often takes "shadow" genes as positive predictions.

As one can see from table 2, the sensitivity of our method is lower in all cases, comparing to the GLIMMER gene-finder, having significatly better specificity. Using annotation of E.Coli, we found that from 228 genes predicted by GLIMMER, and not predicted by our method, 121 are annotated as predicted only by computational methods, 11 ribosomal genes and 12 transposases. From 24 genes predicted by our method and not predicted by GLIMMER, 17 are annotated as predicted only by computational methods and 3 as ribosomal genes. It is not surprising; it is known that ribosomal genes, some other highly expressed genes as well as horizontally transferred genes (the percentage of which is estimated as 10% from the overall number, [Medigue, 1991]) can have rather different (with respect to the average) codon usage, for example, strongly translationally biased codon usage in the case of the ribosomes. It is known also, that preliminary clusterization of genes can enhance existing gene-finding procedures [Mathe et al., 1999,2000].

### *Window-size dependence*

Figure 4 presents our study of window-size dependence of the algorithm effectiveness for two genomes. One can see that the minimal window length, which can be used for the algorithm, is about 100 bp. This value is often characterized as a barrier for all gene-prediction methods based on the analysis of compositional differences. Then, the sensitivity of the algorithm drops monotonically, and, after window size of 400-500 bp, becomes poor.

### *Information content in the triplet distributions*

In this section we study the information content of the triplet distributions in the genetic texts. The question is: what are the contributions to the total amount of information of the triplet distribution, how big are the position-specific information, the contribution connected with correlations between nucleotides and so on? For this purpose we introduce the notion of *mean-field* (or *context free*) approximation of the triplet distributions in the following way:

$$mf_{ijk} = p_i^{(1)} p_j^{(2)} p_k^{(3)},\ p_i^{(1)} = \sum_{jk} f_{ijk}\ ,\ p_j^{(2)} = \sum_{ik} f_{ijk}\ ,\ p_k^{(3)} = \sum_{ij} f_{ijk}\ ,$$

i.e. the mean-field approximation is the distribution constructed from the real triplet distribution neglecting all possible correlations in the order of nucleotides. The $p_i^{(k)}$ are the frequencies of the $i$th nucleotide ($i \in \{A,C,G,T\}$) at the $k$th position of a codon ($k$ = 1..3). In this way we model the 64 frequencies of the real triplet distribution using only 12 frequencies of the three position-specific nucleotide distributions.

It is easy to understand that the phase-shift for $mf_{ijk}$ only rotates the upper (position) indexes:

$$P^{(1)}mf_{ijk} = p_i^{(2)}p_j^{(3)}p_k^{(1)},\ P^{(2)}mf_{ijk} = \left(P^{(1)}\right)^2 mf_{ijk} = p_i^{(3)}p_j^{(1)}p_k^{(2)}.$$

Also it is worth noticing that, applying the $P^{(1)}$ (or $P^{(2)}$) operator several times to the real triplet distribution we will quickly come upon the ( $p_i^{(1)}p_j^{(2)}p_k^{(3)}$, $p_i^{(2)}p_j^{(3)}p_k^{(1)}$, $p_i^{(3)}p_j^{(1)}p_k^{(2)}$ ) triangle:

$$\left(P^{(1)}\right)^3 f_{ijk} = mf_{ijk}.$$

The center of this triangle is close to the point

$$m_{ijk} = p_i p_j p_k,$$

where the $p_i = \frac{1}{3}\left(p_i^{(1)} + p_i^{(2)} + p_i^{(3)}\right)$ is the frequency of the $i$th nucleotide ($i \in \{A,C,G,T\}$). The $m_{ijk}$ distribution is completely randomized distribution (no correlations, no codon position-specific information) and, therefore, has the highest entropy among all the distributions under our consideration. And, this is the hypothetical center of the triplet distributions from all non-coding regions.

Let us consider also the averaged three-phase distribution:

$$f_{ijk}^{av} = \frac{1}{3}\left(f_{ijk} + f_{ijk}^{(1)} + f_{ijk}^{(2)}\right).$$

In the $f_{ijk}^{av}$ distribution all position-specific information is eliminated but it still contains some information about the correlations in the order of nucleotides.

One can measure the distance between two distributions $g_{ijk}$ and $h_{ijk}$ as the relative information of the distribution $g_{ijk}$ with respect to $h_{ijk}$ using the Kullback distance:

$$D(g_{ijk}, h_{ijk}) = \sum_{ijk} g_{ijk} \ln \frac{g_{ijk}}{h_{ijk}}.$$

For our purposes we will use a symmetrized version of the Kullback distance

$$D^{SYM}(g_{ijk}, h_{ijk}) = \frac{1}{2}\left(D(g_{ijk}, h_{ijk}) + D(h_{ijk}, g_{ijk})\right).$$

To visualize the structure of pair-wise distances between different distributions, we use classical metric multidimensional scaling (MDS) (for reference, see, for example, [Torgerson]). The idea of the MDS method is to put the points onto the 2D plane in such a way that to preserve the structure of the pair-wise distances between the points, given by a distance matrix. The resulting pictures are given on the figure 5. The axes of the MDS plot correspond to fictive “principal” coordinates that are assigned to the points to preserve the distances between them. Since shift and rotation of the scatters do not change the distances, we use such a shift that the $m$ point (the highest entropy) is in the (0,0) point of the plot and the rotation angle such that the $f_{ijk}$ (the $f$ point on the plot) is on the negative side of the $x$-axis.

We connect points $f_{ijk}$, $f^{(1)}_{ijk}$, $f^{(2)}_{ijk}$ by solid line. It is the “three-phases” triangle, corresponding to the real triplet distributions in the correct, first and second phases respectively. The second, dashed triangle connects the points of the mean-field approximation ($mf_{ijk}$, $P^{(1)}mf_{ijk}$, $P^{(2)}mf_{ijk}$). In the table 3 we also present the distance matrix calculated for the triplet distributions in *H.pylori*.

Let us discuss the general features of the pictures. Qualitatively, the information content (relative entropy) of a point on the plots in fig.5 is proportional to the distance from the center of plot (0,0). The maximum of information is contained, of course, in the $f_{ijk}$ distribution (the $f$ point), which is the most distant point on the plots. For example, in the case of *H.pylori*, the relative information of the tripet distribution equals 0.29. The value is higher in the case of *C. crescentus* (0.39) and less in the case of *S.cerevisiae* genome (0.18). In fact, high information content of the triplet distribution in the correct phase gives more contrast cluster structure and better quality of unsupervised gene recogniton.

The distances $D^{SYM}(f_{ijk}, f^{(1)}_{ijk})$ and $D^{SYM}(f_{ijk}, f^{(2)}_{ijk})$ are approximately equal (0.46 and 0.44, for *H.pylori*) and bigger than the distance between $f^{(1)}_{ijk}$ and $f^{(2)}_{ijk}$ (0.32 for *H.pylori*). This can be explained if the correlations in the order of codons in the coding sequences are small (our study shows that this is the case for, at least, bacterial and yeast genomes). If so, then the distributions in the first and second phases can be reconstructed from the $f_{ijk}$ using only position-specific frequencies of nucleotides and di-nucleotides. Indeed, the information contents of $f^{(1)}_{ijk}$ and $f^{(2)}_{ijk}$ are less than in $f_{ijk}$. (0.19 and 0.19 against 0.29, for *H.pylori*).

Shifted distributions are reconstructed from the initial distribution, applying phase-shifts operators $P^{(1)}$ and $P^{(2)}$. In all cases these reconstructions (points *P1f* and *P2f*), calculated using assumption about smallness of correlations in the order of codons, are very close to the real distributions in the first and second phases (points *f1* and *f2* on the plots).

The "mean-field approximation" triangle is isosceles with it's center approximately in the $m_{ijk}$ point. The difference in sizes of the "three-phases" triangle and the "mean-field approximation" triangle reflects presence of correlations in the order of nucleotides. In fig.5, this difference is small in the case of *C. crescentus* and considerable in other three genomes. From the table 3, one can see that in the case of *H.pylori,* the average length of the "three-phases" triangle side is 0.41 while the same value for the "mean-field approximation" triangle is only 0.16. The loss of information after neglecting all correlations in the order of nucleotides (the distance from *f* to *mf* points) is 0.21 in the case of *H.pylori* and 0.15 in the case *C. crescentus.*

## Implementation

All datasets were prepared from sequences in the GenBank flat-file format. The programs used for data analysis, including simple implementation of the K-means clusterization algorithm, were written in Java and are available with instructions at the accompanying web page: http://www.ihes.fr/~zinovyev/bullet/. These programs actively use the BioJava programming package. Technically, the data visualization and all illustrations were produced using the ViDaExpert data visualization tool under Windows, and are available at the supplementary web-page. For producing the MDS plots we used a procedure of classical metric multidimensional scaling from the Matlab environment.

## Discussion

Seven clusters structure of oligomer distributions in genetic texts plays important role in ability of modern gene-finders for unsupervised (and, to lesser extent, also for supervised) learning in prokaryotic genomes. Actually existence of the structure makes the prokaryotic gene-finding so efficient. Using seven hidden states for hidden Markov modeling approach in gene-prediction was introduced long ago (see, for example, Borodovsky, 1993). Though being widely exploited, this structure was never visually presented and analyzed by pure data exploratory study means.

Our study shows relatively high performance of using only short n-mers, like triplets. It means that an essential part of the information needed to discriminate between coding and non-coding regions is already contained in their triplet distributions. Using hexamer frequencies (that is common practice in modern gene-finders) can be more sensitive, but also can lead to certain undesirable effects. One needs more sequence information to evaluate hexamer frequencies, and, as a result, this fact can lead to "overfitting" effects, leading to worse specificity. We demonstrated this fact, using visual analysis of positive predictions of GLIMMER gene-finder.

We demonstrated as well that the correlations in the order of codons are small with respect to the interphase and "coding-noncoding" distance. This fact is interesting by itself, and is not completely trivial. In particular, it implies that in commonly used for gene prediction seven-states HMMs, the weights of different (coding) states are not at all independent: their dependence in the case of HMMs of order 2, in zero approximation is given by the formulas in the beginning of the "Methods" section. Another somehow unexpected observation is that the sizes of clusters in the phase triangle are similar. It would be natural to expect the cluster which corresponds to the "correct" P0 (or C0) phase to be more compact then P1 (C1) and P2 (C2) clusters, but this is not the case.

From the constructed representations of datasets it is clear that the spatial structure of triplet distributions is almost completely determined by two factors: 1) the frequency distribution of the 64 codons in the coding phase; 2) the dispersion of the codon frequency distribution. The latter one is related to the structure of codon usage over all genes in a genome, which is known to be inhomogeneous (see, for example, Medigue, 1991), especially in such fast-growing organisms as E.Coli and B.Subtilis, where the translational bias shapes the codon usage differently for different groups of genes. Nevertheless, the dispersion is smaller with respect to the phase-phase and coding-noncoding distances, which makes the gene-prediction possible even without preliminary genes classification.

From the figures, it is evident that the distribution structure renders linear discrimination analysis (sometimes applied in this situation) inapplicable. Applying linear methods in this case would lead to the incorrect conclusion that the dataset is not well-separable and that this measure is less effective than others with respect to linear discrimination function. For example, in the case of Helicobacter pylori, linear discrimination yields a specificity of ~0.83 (which means many false positives), while the method we proposed yields ~0.97. This fact stresses that understanding the spatial structure of a learning dataset is necessary for reasonable applications of pattern recognition methods.

Frequency normalization plays an important role in cluster structure formation. It indicates the importance in distinguishing coding and non-coding regions of those codons that may not have high frequency values but considerably change their frequencies after phase-shift (codons that are "prohibited,", due to codon bias.)

Basically, distribution of non-overlapping triplets that is efficient for gene recognition corresponds to a high value of mutual information in three consecutive letters, i.e.,

$$M = \sum_{ijk} f_{ijk} \log_2 \frac{f_{ijk}}{p_i\, p_j\, p_k},$$

where $p_i$ is the average frequency of the *i*th nucleotide $i \in \{A, C, G, T\}$ This value may be zero only in the case $f_{ijk} = p_i\, p_j\, p_k$. In this case, we would have $P^{(1)} f_{ijk} = P^{(2)} f_{ijk} = f_{ijk}$, *i.e.,* phase-shift does not change the codon distribution. High values of *M* guarantee the presence of a “three-phase triangle” in the data space, as well as the formation of a cluster structure.

In this paper, using visual analysis of spatial dataset structure and very simple clustering technique, we have shown that using learning dataset is not necessary in order to accurately solve gene recognition tasks, at least in that DNA segments with high concentrations of coding information (compact genomes). This property is very useful, since the problem of choosing a "good" learning dataset is not very well defined (see, for example, [Mathe, Sagot et al.]).

The method proposed can be applied for the rough annotation of unassembled genomes, since it does not require preliminary extraction of ORFs. This makes it useful for inexpensive genome survey projects. Also it allows efficient analysis of performance of existing prokaryotic gene-finders. One more (and not the least) advantage of the visual representation of oligomer distributions is that it facilitates understanding of the subject by those who just enter this field.

**Acknowledgements**

The authors thank Alessandra Carbone (IHES, France) for very fruitful discussions, Misha Gromov for the interest he expressed in this work, Noah Hardy and Arndt Benecke for editing the manuscript.

Table 1

Summary table of results for assessing the method on the nucleotide level

| Sequence | $L$ | $W$ | $p$ | $\nu^{(3)}$ | % of coding bases | CP | CC | $Sn_1$ | $Sp_1$ | $Sn_2$ | $Sp_2$ |
|---|---|---|---|---|---|---|---|---|---|---|---|
| *Helicobacter pylori* | 1643831 | 300 | 120 | 0.35 | 90 | 0.68 | 0.28 | 0.93 | 0.97 | 0.93 | 0.98 |
| *Caulobacter crescentus* | 4016947 | 300 | 300 | 0.21 | 91 | 1.07 | 0.16 | 0.93 | 0.97 | 0.94 | 0.98 |
| *Prototheca wickerhamii* | 55328 | 120 | 18 | 0.17 | 49 | 0.83 | 0.11 | 0.82 | 0.93 | 0.84 | 0.95 |
| *Saccharomyces cerevisiae* chromosome III | 316613 | 399 | 99 | 0.16 | 69 | 0.45 | 0.10 | 0.90 | 0.88 | 0.90 | 0.90 |
| *Saccharomyces cerevisiae* chromosome IV | 1531929 | 399 | 120 | 0.15 | 73 | 0.48 | 0.09 | 0.89 | 0.91 | 0.92 | 0.92 |

Table 2

Comparing the method with GLIMMER gene-predictor

| Sequence | CLUSTER | | GLIMMER | |
|---|---|---|---|---|
| | Sn | Sp | Sn | Sp |
| *Helicobacter pylori* | 0.94 | 0.95 | 0.96 | 0.78 |
| *Haemophilus influenza* | 0.93 | 0.88 | 0.96 | 0.84 |
| *Escherichia coli* | 0.91 | 0.87 | 0.96 | 0.76 |
| *Bacillus subtilis* | 0.89 | 0.95 | 0.97 | 0.79 |
| *Caulobacter crescentus* | 0.89 | 0.76 | 0.94 | 0.60 |

Table 3

Symmetric Kullback distances between triplet distributions for *Helicobacter pylori*

| | $f$ | $f^{(1)}$ | $f^{(2)}$ | $f^{av}$ | $P^{(1)}f$ | $P^{(2)}f$ | $mf$ | $m$ | $P^{(1)}mf$ | $P^{(2)}mf$ |
|---|---|---|---|---|---|---|---|---|---|---|
| $f$ | | 0.46 | 0.44 | 0.18 | 0.44 | 0.45 | 0.21 | 0.29 | 0.41 | 0.42 |
| $f^{(1)}$ | | | 0.32 | 0.12 | 0.06 | 0.30 | 0.32 | 0.19 | 0.12 | 0.31 |
| $f^{(2)}$ | | | | 0.11 | 0.35 | 0.06 | 0.31 | 0.19 | 0.32 | 0.13 |
| $f^{av}$ | | | | | 0.14 | 0.12 | 0.14 | 0.08 | 0.14 | 0.14 |
| $P^{(1)}f$ | | | | | | 0.31 | 0.25 | 0.13 | 0.06 | 0.24 |
| $P^{(2)}f$ | | | | | | | 0.23 | 0.12 | 0.24 | 0.06 |
| $mf$ | | | | | | | | 0.05 | 0.16 | 0.16 |
| $m$ | | | | | | | | | 0.05 | 0.05 |
| $P^{(1)}mf$ | | | | | | | | | | 0.16 |
| $P^{(2)}mf$ | | | | | | | | | | |

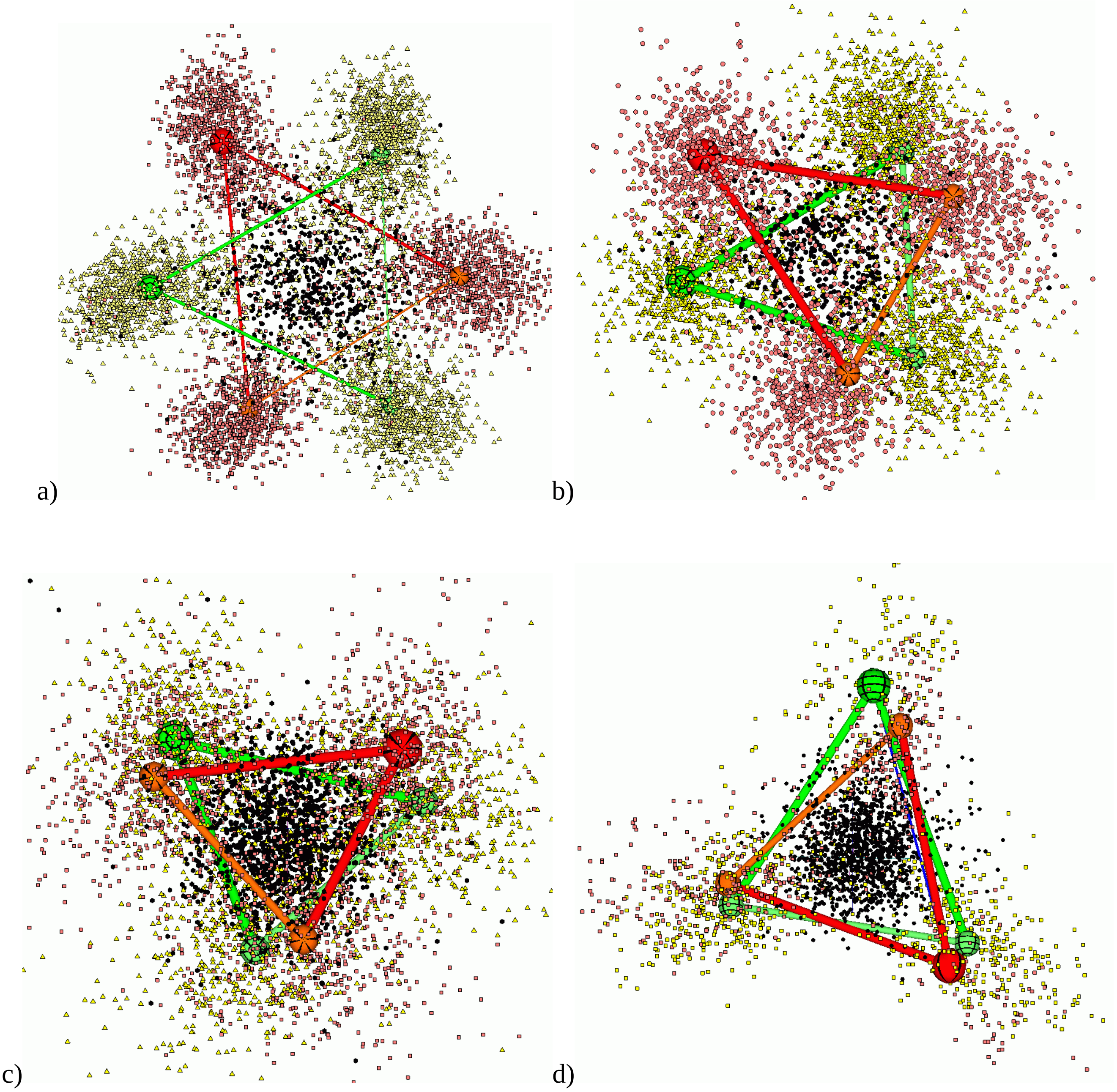

● – non-coding, ■ – exons in forward strand, ▲ – exons in complementary strand ,

– $f_{ijk}$, – $P^{(1)} f_{ijk}, P^{(2)} f_{ijk}$ ; – $\hat{f}_{ijk}$ , – $P^{(1)} \hat{f}_{ijk}, P^{(2)} \hat{f}_{ijk}$

a) *Caulobacter crescentus* (GenBank NC_002696);
b) *Helicobacter pylori* (GenBank NC_000921);
c) *Saccharomyces cerevisiae* chromosome IV (GenBank NC_001136);
d) *Prototheca wickerhamii* mitochondrion (GenBank NC_001613).

Figure 1. Visualization of genetic sequences in the space of triplet frequencies

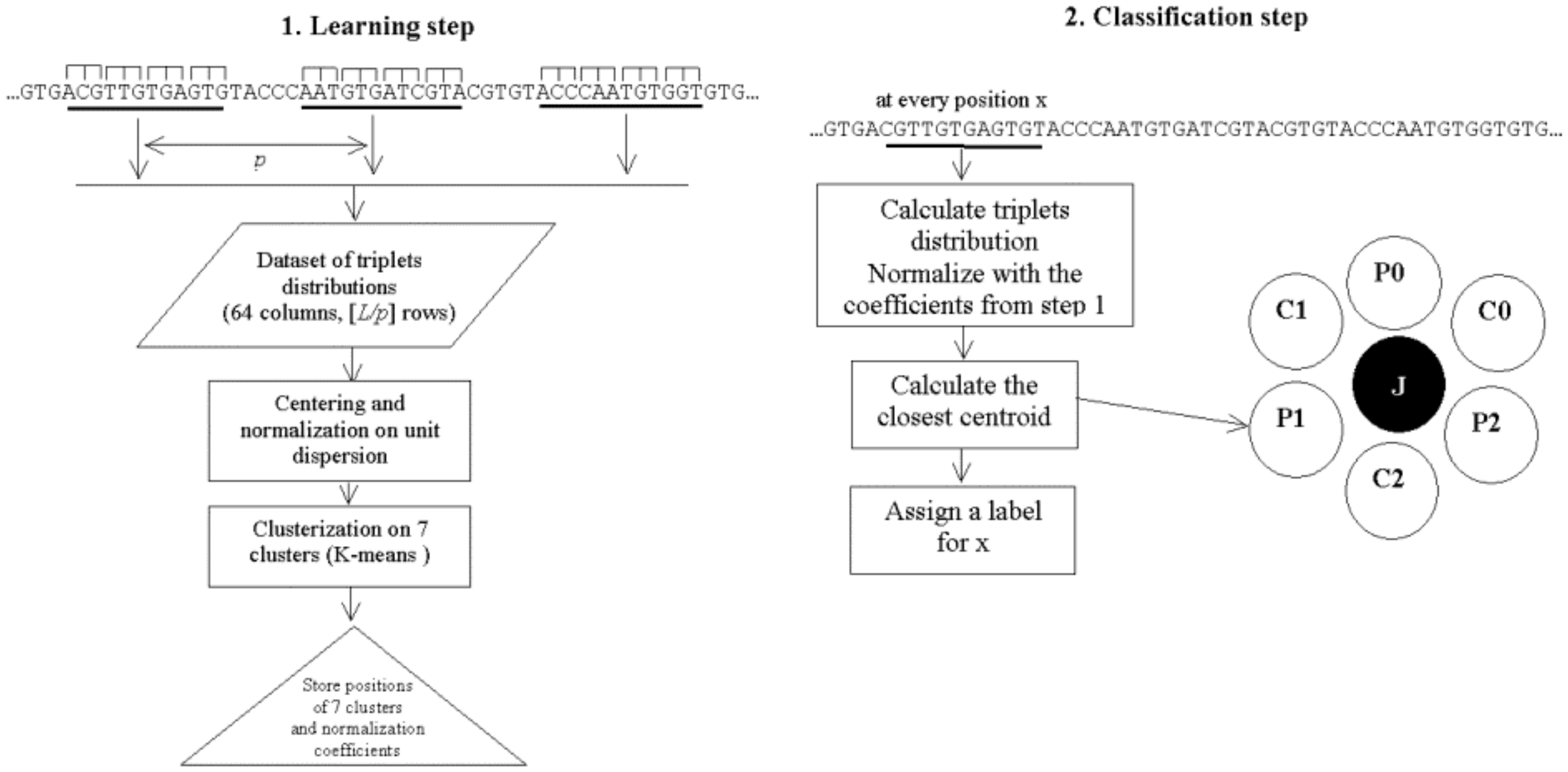

Figure 2. Graphical representation of the method.

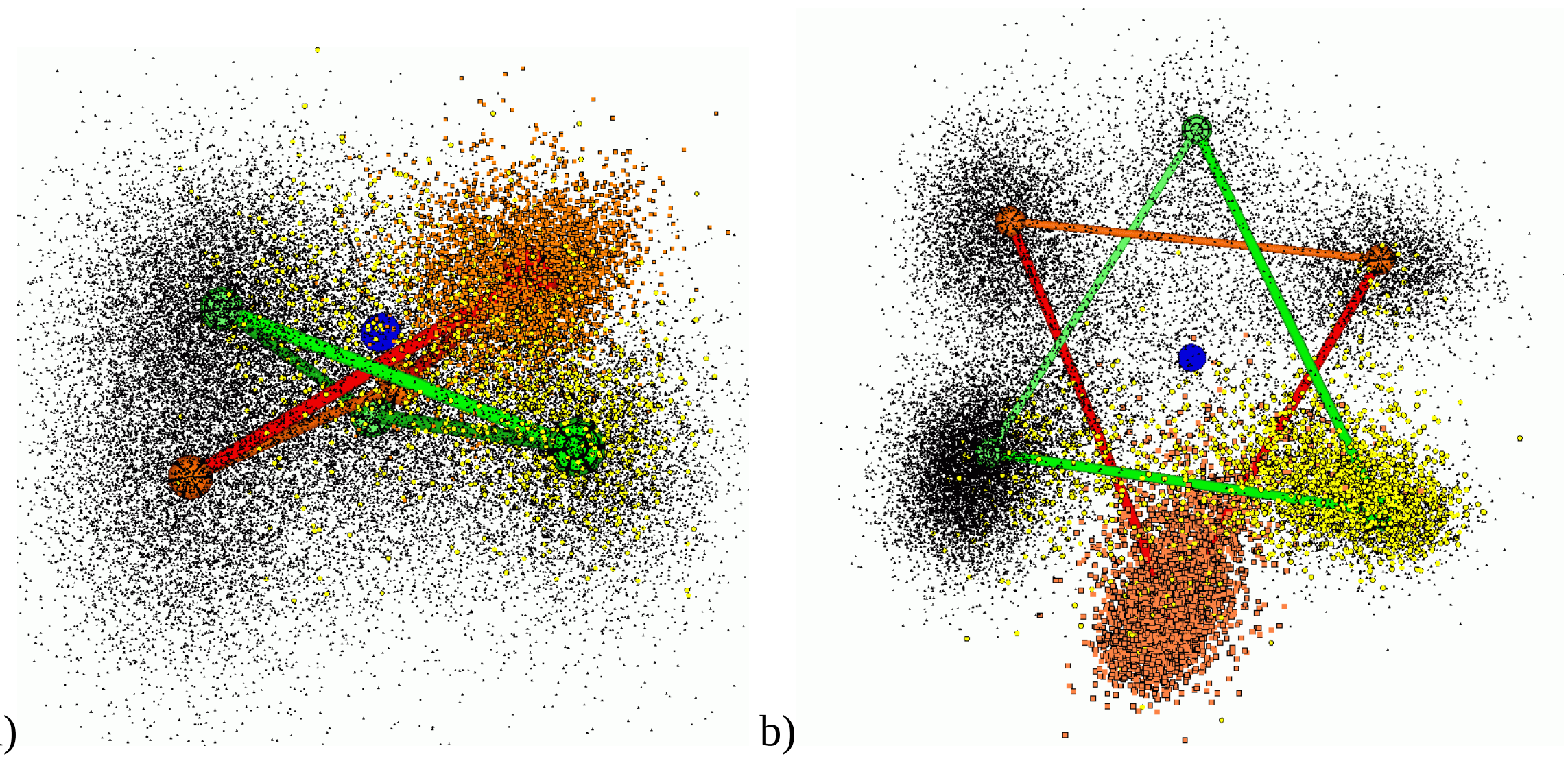

– negative predictions, – true positive predictions, – false positive predictions,
– $f_{ijk}$, – $P^{(1)} f_{ijk}, P^{(2)} f_{ijk}$; – $\hat{f}_{ijk}$, – $P^{(1)} \hat{f}_{ijk}, P^{(2)} \hat{f}_{ijk}$

Figure 3. Visualization of the distribution of predictions of GLIMMER gene-finder in 64-dimensional space of codon frequencies.
Every point corresponds to one ORF. Red and green triangles denote the same structures as described at the figure 1.

a) *Escherichia coli*. Projection on the 1st and 3d principal components.
b) *Caulobacter crescentus*. Projection on the 1st and 2d principal components.

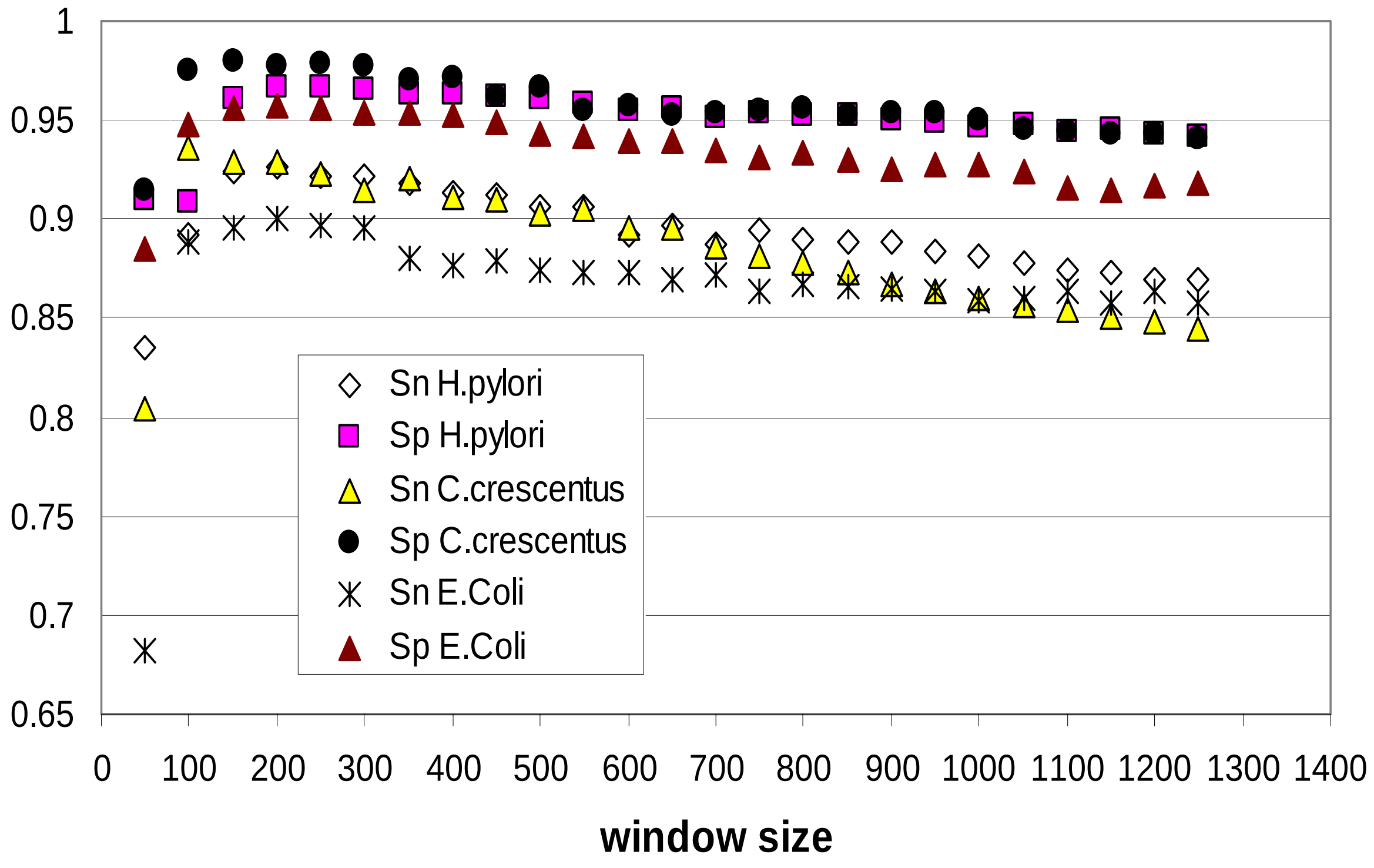

Figure 4. Window-size dependence of the algorithm

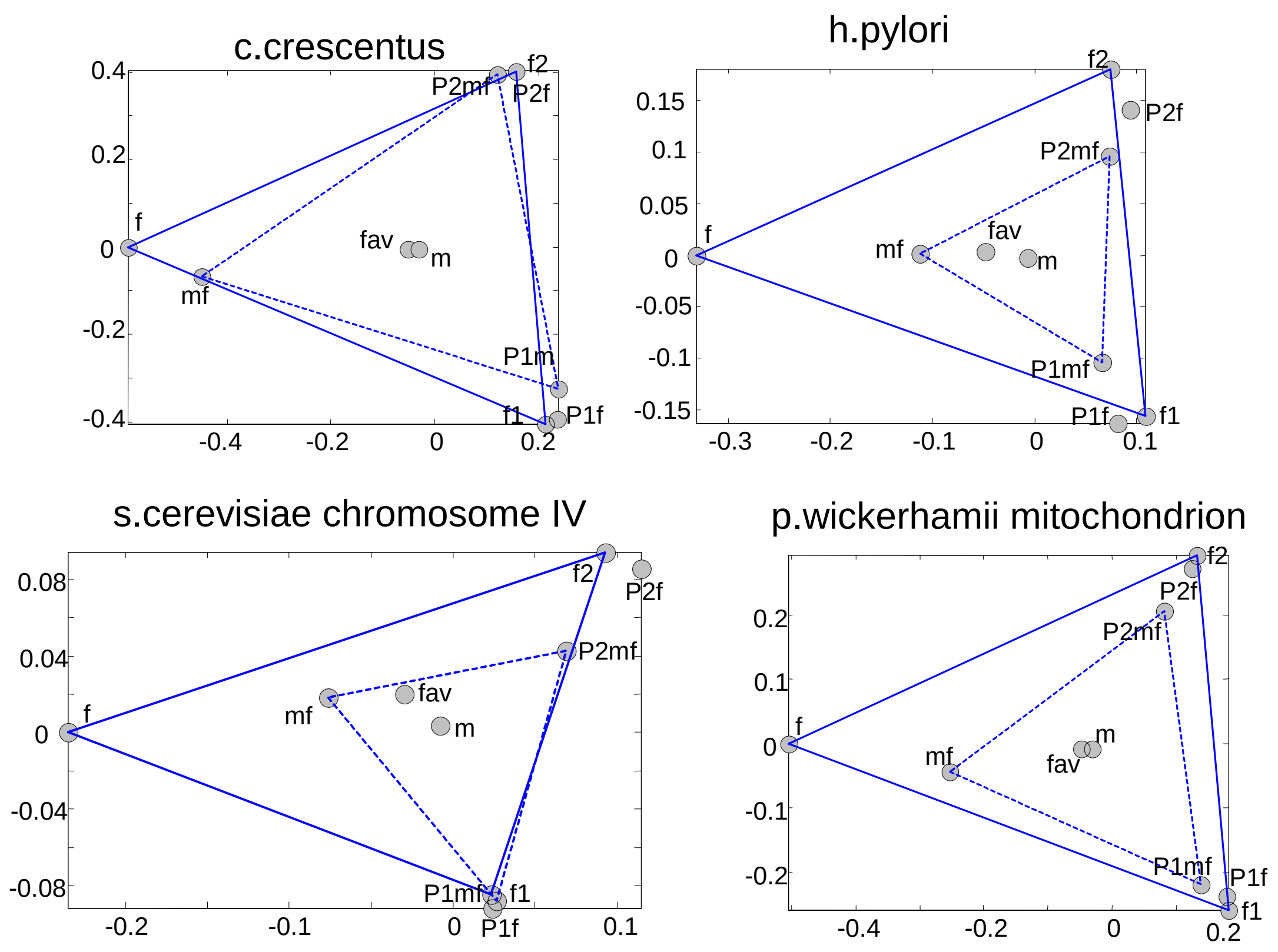

Figure 5. MDS plots (see explanation in the text) representing the structure of Kullback distances between different distributions. The solid triangle is the “three-phase” triangle, calculated from the real gene sequences. Dashed triangle is the corresponding “mean-field” (context free) approximation.

Annotations:

$f = f_{ijk}$ – real triplet distribution in the correct phase;

$f1 = f_{ijk}^{(1)}$ – real triplet distribution in the first phase;

$f2 = f_{ijk}^{(2)}$ – real triplet distribution in the second phase;

$P1f = P^{(1)} f_{ijk}$ – calculated distribution in the first phase;

$P2f = P^{(2)} f_{ijk}$ – calculated distribution in the second phase;

$fav = \frac{1}{3}\left(f_{ijk} + f_{ijk}^{(1)} + f_{ijk}^{(2)}\right)$ – average distribution of triplets;

$mf = p_i^{(1)} p_j^{(2)} p_k^{(3)}$ – the mean-field (context free) approximation of the codon usage, $p_i^{(k)}$ – are the frequencies of the $i$th nucleotide ($i \in \{A,C,G,T\}$) at the $k$th position of codon ($k = 1..3$).

$P1mf = P^{(1)} mf = p_i^{(2)} p_j^{(3)} p_k^{(1)}$ – mean-field approximation in the first (shifted) phase;

$P2mf = P^{(2)} mf = p_i^{(3)} p_j^{(1)} p_k^{(2)}$ – mean-field approximation in the second (shifted) phase;

$m = p_i \, p_j \, p_k$ – randomized distribution (the highest entropy).